\documentclass[submission, Phys]{SciPost}

\usepackage{amsmath}
\usepackage{amssymb}
\usepackage{amsbsy}
\usepackage{graphicx}
\usepackage{color}
\usepackage{lmodern}

\usepackage{enumerate}
\usepackage{mathtools}
\usepackage{ulem}

\def\be{\begin{equation}}
\def\ee{\end{equation}}
\def\bfi{\begin{figure}}      
\def\efi{\end{figure}}
\def\bea{\begin{eqnarray}}
\def\eea{\end{eqnarray}}

\begin{document}

\begin{center}{\Large \textbf{An objective criterion for cluster detection in stochastic epidemic models}}\end{center}

\begin{center}
E. Lippiello\textsuperscript{1},
P. Bountzis\textsuperscript{2}
\end{center}

\begin{center}
{\bf 1}Department of Mathematics and Physics, University of Campania ``Luigi Vanvitelli'', 81100, Caserta, Italy
\\
  {\bf 2}Department of Geophysics, School of Geology, Aristotle University of Thessaloniki, GR54124, Thessaloniki, Greece
\\
* eugenio.lippiello@unicampania.it
\end{center}

\begin{center}
\today
\end{center}


\section*{Abstract}
{\bf   The correct identification of clusters is crucial for an accurate monitoring of the spread of a disease and also in many other natural, social and physical phenomena which exhibit an epidemic structure. Nevertheless, even when an accurate mathematical model is available, no simple tool exists which allows one to identify how many independent clusters are present and to link elements to the appropriate clusters.
Here we develop an automatic method for the detection of the internal structure of the clusters and their number, independently of the model that describes the dynamics of the phenomenon. It is substantially based on the difference of the log-likelihood $\delta {\cal LL}$\, that is evaluated when all elements are connected and when they are grouped into clusters. As a function of the number of connected elements $\delta {\cal LL}$ presents a change of slope and a singularity which can be both used in cluster identification. Our method is validated for an epidemic model with a minimal temporal structure and for the Epidemic Type Aftershock Sequence model describing the spatio-temporal clustering of earthquakes. }

\noindent\rule{\textwidth}{1pt}
\tableofcontents\thispagestyle{fancy}
\noindent\rule{\textwidth}{1pt}
\vspace{10pt}

\section{Introduction}

Epidemic models provide an accurate mathematical description of many natural and social processes with many fruitful applications in biology, computer and social science as well as chemical, physical and geophysical systems \cite{Har63,Ath72,Jag75,CG85}.
The common situation concerns a data set containing $N$ elements which are distributed in time and space and can be viewed as a network composed of $N$ nodes and zero links. The central task is to create links only between correlated elements.
In the case of a disease spreading, for instance, the elements are individuals who have been infected at a given time and in a given place, and links must be added to the correct infector-infectee pairs. In seismic occurrence, within a point-process description, the elements are earthquakes and the task corresponds to the establishment of links between the triggering and the triggered subsequent earthquakes due to the released stress.
A cluster or a community, within this epidemic framework, corresponds to a dendogram containing all elements which originate from a common ancestor, i.e. the patient zero or index case in epidemiology or the mainshock in seismology.   
The identification of communities can be framed into a general issue, known as cluster analysis,  which is often considered  a branch of pattern recognition and artificial intelligence \cite{KR90,FH16}. Within this direction it becomes fundamental to establish an automatic procedure for the classification of similar objects into clusters.

All epidemic processes present a common mathematical description originating from a renewal equation which allows one to express the occurrence probability $\lambda(t_i,x_i)$ of a new element $i$ at time $t_i$, in position $x_i$, in terms of the history of all previous elements with occurrence times $t_j<t_i$, in a given target region $\Sigma$, 
\be
\lambda(t_i,x_i)=\sum_{j |t_j<t_i}p_{ij} +\mu.
\label{lambda}
\ee
Here, $p_{ij}$ is the probability the $j$-th element to trigger the subsequent element $i$, or equivalently in epidemiology, to induce  the subsequent infection of element $i$. The term $\mu$ in Eq.(\ref{lambda}) represents the rate of imported cases from outside $\Sigma$, i.e the rate of ancestors or immigrants.
Our aim is to use the information contained in $p_{ij}$ to identify links between true correlated elements. This procedure corresponds to the construction of an adjacency matrix $Q$, containing $N \times N$ elements and presenting a block structure with elements $q_{ij}=0$ if $i$ and $j$ belong to different clusters and $q_{ij}=1$ otherwise. The number of non-null elements of the matrix $Q$ corresponds to the total number of links, $n_{link}$, in the epidemic network. 
We observe that, since each cluster originates from a single  immigrant, their number coincides with the number of independent clusters or, equivalently, with  the number of blocks in the matrix $Q$.

In this study, we present a novel procedure which is substantially based on the log-likelihood difference $\delta {\cal LL}=\sum_{i=1}^N \delta {\cal LL}_i$ with
\be
\delta {\cal LL}_i=
\log\left(\sum_{j |t_j<t_i}p_{ij} +\mu \right)-
\log\left(\sum_{j |t_j<t_i}p_{ij}q_{ij} +\mu' \right),
\label{ll1}
\ee
between a process where links exist for all pair of elements (first sum in Eq. \ref{ll1}) and one where only pairs with $q_{ij}=1$ are considered
\footnote{In both definitions of the log-likelihood an additive term is present. This term comes from the normalization and, being equal to $N$, is cancelled by the subtraction.}.
We add the term $\mu'$ to avoid the logarithmic divergence and we set $\mu' \ll \mu$. 
We show that $\delta {\cal LL}$ presents two distinct singular behaviors as a function of the links number, which we define $n_1^*$ and $n_2^*$. When $n_{link}=n_1^*$ the matrix $Q$ contains the optimal number of blocks corresponding to the true number of immigrants in the process, whereas $n_{link}=n_2^* \ge n_1^*$ is the minimal number of links such as $q_{ij} =1$ for all pairs of correlated elements.
We test our method for two models based on different expressions of $p_{ij}$ in Eq.(\ref{lambda}).

\section{The method}


We consider a data set containing $N$ elements grouped in $K$ clusters,  each one corresponding to a different immigrant, and we define $[{\cal X}]=\{ {\cal X }_k \}_{k=1,...,K}$ as the ``a-priori'' partition which is unknown in a real data set. Our task 
is to extract from the spatio-temporal organization of the $N$ elements, an ``a-posteriori'' partition $ [{\cal Y}]$ which represents the best approximation of $[{\cal X}]$.

\begin{figure*}
  \includegraphics[width=16cm,height=8cm]{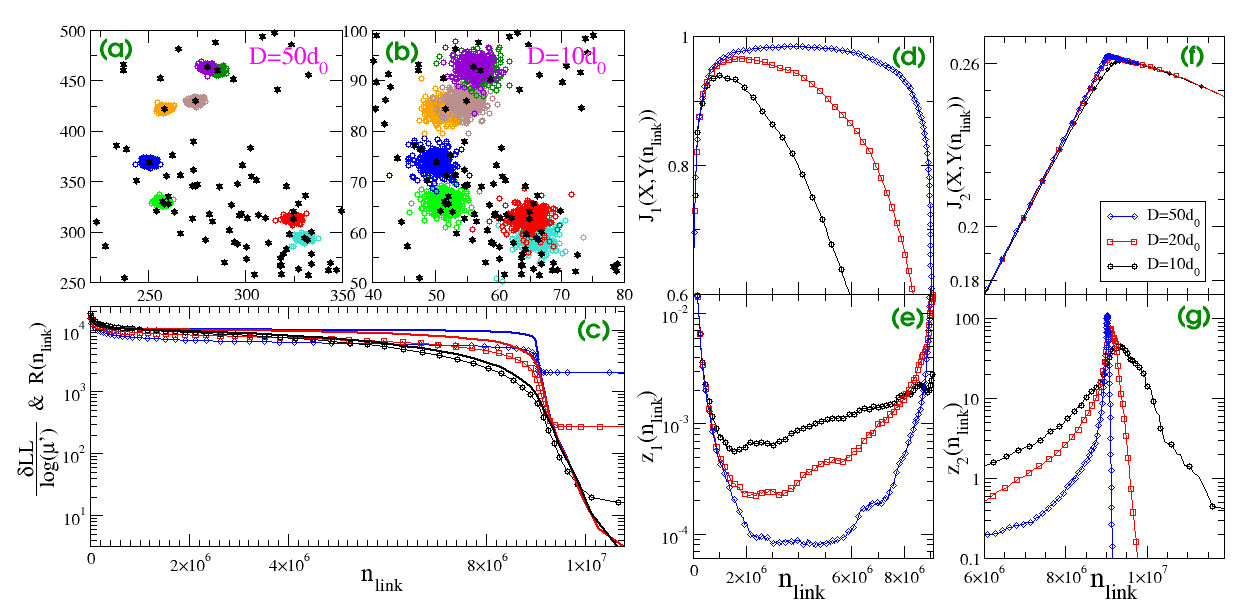}
  \caption{The clustering model with $D=50d_0$ (a) and $D=10d_0$ (b). Same colors and symbols are used for elements of the same cluster. Immigrant elements are indicated by black stars. (c)    Continuous lines represent $R\left(n_{link}\right)$ whereas different symbols are used for  $\delta{\cal LL}\left(n_{link}\right)$ for different values of $D/d_0$:  blue diamonds for $D=50d_0$, red squares for $D=20d_0$ and black circles for $D=10d_0$.
    The same color codes and symbols are applied for the index $J_1({\cal X},{\cal Y}(n_{link}) )$ (d) the slope $z_1(n_{link})$ (e) the Jaccard index  $J_2({\cal X},{\cal Y}(n_{link}))$ (f) and the effective exponent $z_2(n_{link})$ (g).}
  \label{fig1}
\end{figure*}

We characterize each partition $[{\cal Y}\left(n_{link}\right)]$ according to the number of non-null elements in its associated adjacency matrix, $Q({\cal Y})$ with $n_{link}$ 
elements equal to one ($q_{ij}=1$)  if $i$ and $j$ belong to the same cluster, whereas all other elements are null. We set to zero the diagonal elements $q_{ii}=0$ and assume that events are sorted according to their temporal occurrence, i.e. $t_i>t_j$ for $i>j$, which leads to a lower triangular matrix $Q({\cal Y})$.

Our first task is the identification of a partition $[{\cal Y}\left(n_{link}\right)]$ that is composed by a number of immigrants $R$ which best overlaps with the $K$ immigrants present in the partition $[{\cal X}]$. To this extent, starting from a partition $[{\cal Y}\left(n_{link}\right)]$ we add $\delta n$ non null matrix elements $q_{ij}$ to $Q({\cal Y}\left(n_{link}\right))$ obtaining the final partition $[{\cal Y}\left(n_{link}+\delta n\right)]$. 
By construction, the number of immigrants $R\left(n_{link}\right)$ in the starting partition is larger  or equal to the one in the new partition
$R\left(n_{link}\right) \ge R\left(n_{link}+\delta n\right)$. We first observe that  an element $i$ is assigned as an immigrant if and only if $q_{ij}=0$ for all elements $j<i$. We next indicate with $\rho$ the probability that a new added matrix element reduces the number of immigrants, i.e. $R\left(n_{link}+1\right) = R\left(n_{link}\right)-1 $.
 For $n_{link} \ll K$ any element that is added to $Q$ is  expected to reduce $R$ so  $\rho \simeq 1$. Increasing $n_{link}$, $\rho$ is expected to decrease up to  its minimum value at  $n_{link}=n_1^*$ when it is very unlikely an immigrant to be associated with a previous element.  
The minimum of $\rho$ therefore identifies the partition  $[{\cal Y}\left(n_1^*\right)]$  with the optimal number of immigrants $R\left(n_1^*\right) \simeq K$. We finally observe that
the probability $\rho$ can be obtained from the derivative $z_1\left(n_{link}\right)=-\frac{\partial R\left(n_{link}\right)}{\partial n_{link}}$, which is easily computed from the sample. Indeed, 
assuming that $\delta n$ is sufficiently small such that $\rho$ is almost constant for all $R\left(n_{link}+j\right)$, with $0\le j \le \delta n$, the expected value of  $R$ is equal to
$E[R]= \sum_ {j=0}^{\delta n} 
{\delta n \choose j}  (1-\rho)^{\delta n-j}\rho^j\left (R\left(n_{link}\right)-j\right)=R\left(n_{link}\right)-\delta n \rho$ and
therefore $z_1\left(n_{link}\right)\simeq \rho$.

We remark that even if the  partition $[{\cal Y}\left(n_1^*\right)]$ contains a number of clusters $R\left(n_1^*\right)$ which well approximates $K$, we expect that $n_1^*$ is significantly smaller than the number of links present in the partition $[{\cal X}]$, since triggered/infected elements can produce their own descendants/infections, and the adjacency matrix $Q({\cal Y})$ contains too many null elements. For this reason, in the following we introduce another criterion for the identification of the optimal number of links $n_2^*>n_1^*$ in the partition $[{\cal Y}]$.
We start from the log-likelihood difference $\delta {\cal LL}_i$ defined in Eq.\ref{ll1} and we focus on the elements $i$ identified as immigrants. For these elements, $q_{ij}=0$ $\forall j$ and therefore $\delta {\cal LL}_i=-\log(\mu')+\log\left(\sum_{j |t_j<t_i}p_{ij} +\mu \right)$. Under our assumption $\mu' \ll \mu<1$, the last term is always much smaller than $\log(\mu')$ and it can be neglected.
Furthermore, for the non-immigrant elements, since $p_{ij}q_{ij}  \gg \mu'$, $\delta {\cal LL}_i \ll -\log(\mu')$ and therefore, for small values of $n_{link}<n_1^*$,
$\delta {\cal LL}=\sum_{i=1}^N\delta {\cal LL}_i$ is dominated by the contribution of the $R(n_{link})$ immigrant elements leading to
\be
\delta {\cal LL} \simeq -R(n_{link}) \log(\mu').
\ee

However, for increasing $n_{link}>n_1^*$, the function $R(n_{link})$ is converging fast  to zero eliminating the contribution of immigrant elements to $\delta {\cal LL}$. For this reason in the following we focus on $\delta {\cal LL}_i$ for non-immigrant  elements and we observe that for any $i$ the sum over the $j<i$ elements in the first logarithm in Eq.(\ref{ll1}) can be split into two contributions: The first consists of the $n(i)$ elements which belong to the same cluster in the partition $[{\cal X}]$ (internal elements with $k(j) =k(i)$) and the second consists of the $i-1-n(i)$ external elements  ($k(j)\ne k(i)$), where $k(i)$
is the index of the cluster to which the element $i$ belongs
in the partition $[{\cal X}]$. More precisely, we indicate with $ \overline{p}$ the average value of $p_{ij}$ over the $n(i)$ internal elements with $k(j)=k(i)$ and with  $\epsilon \overline{p}$ its average value over the remaining $i-1-n(i)$ external ones with $k(j) \ne k(i)$. Accordingly, Eq.(\ref{ll1}) can be written as
\be
\delta {\cal LL}_i=\log\left(n(i) \epsilon \overline p\right)+\log\left(1+\xi_i\right)
-\log\left(\sum_{j |t_j<t_i}p_{ij}q_{ij} +\mu' \right),
\label{ll5}
\ee
where $\xi_i=\left(\mu+(i-1-n(i)) \epsilon \overline{p}\right)/\left(n_i  \overline{p}\right)$.
 We next consider a partition $[{\cal Y}\left (n_{link}\right)]$ which presents $n(i)+\delta n(i)$ elements $j$ with $q_{ij}=1$ in its associated adjacency matrix $Q([{\cal Y}])$ and we consider the case $\delta n(i) \ll n(i)$. Defining $x_i=1+\delta n(i)/n(i)$ ($x_i \simeq 1$) and taking into account that $\mu'$ can be neglected, Eq.(\ref{ll5}) can be written as 
\be
\delta {\cal LL}_i \simeq
\begin{cases} 
  & \log\left(1+ \xi_i\right) - \log\left(x_i\right)\quad  \text{$x_i\le 1$} \\
    & \log\left(1+ \xi_i\right) - \log\left(1+\epsilon (x_i-1)\right)\quad  \text{$x_i>1$}.
\end{cases}
\label{ll6}
\ee
We next evaluate the effective exponent, $z(x_i) = -\frac {\partial \log\left(\delta {\cal LL}_i (x_i) \right)}{\partial \log(x_i)}$ which from Eq.(\ref{ll6}) reads
\be
z(x_i) \simeq 
\begin{cases}
  \frac{1}{\log(1+\xi_i) -\log(x_i)}
  \quad & \text{$x_i\le1$} \\
  \frac{1}{\log(1+\xi_i)- \log(1+\epsilon (x_i-1))} \frac{\epsilon x_i}{1+\epsilon (x_i-1)}
& \quad \text{$x_i>1$}.
\end{cases}
\label{zeta}
\ee
We remark that, since the average value ($\overline{p}$) of $p_{ij}$ among internal elements is usually significantly greater than the average value ($\epsilon \overline{p}$)  with external ones,  $\epsilon$ is expected to be much smaller than $1$.
As a consequence the quantity $z(x_i)$ presents a clear discontinuity at the point $x_i=1$ ($z(1^+) \simeq\epsilon z(1^-)$).
According to our construction the value $x_i=1$ corresponds to the case where the matrix 
$Q([{\cal Y}])$  contains, in the $i$-th column, the same non-null elements $q_{ij}$ present in the matrix $Q([{\cal X}])$.  Therefore, the discontinuity of $z(x_i)$ can be used to identify the
optimal partition, in the sense that ${\cal Y}_{k(i)}$ presents the best overlap with ${\cal X}_{k(i)}$.
The same argument holds for all non-immigrant elements so we expect that
$\delta {\cal LL}\left(n_{link}\right)$ presents a corner point at $n_{link}=n_2^*$ which reflects a discontinuity in the effective exponent $z_2\left(n_{link}\right)$ defined as
\be
z_2\left(n_{link}\right)= -\frac {\partial \log\left(\delta {\cal LL} \left(n_{link}\right)\right)}{\partial \log\left(n_{link}\right)}.
\label{zeta2}
\ee

Summarizing, we introduced two criteria for the identification of the partitions $[{\cal Y}\left(n_1^*\right)]$ and $[{\cal Y}\left(n_2^*\right)]$. The first one corresponds to a minimum of $z_1\left(n_{link}\right)$ and its partition represents the optimal detection of the immigrant elements. The second one corresponds to a discontinuity in $z_2\left(n_{link}\right)$ and its partition gives the optimal internal structure of the clusters. 

\section{Validation method}

In the following we test our method in numerical simulations where we adopt an hierarchical algorithm \cite{dAGL18} for the simulation of a cascading process according to Eq.(\ref{lambda}) .
In numerical simulations the a-priori partition $[{\cal X}]$ is  known.
A very common method to test if an ``a-posteriori''  partition $[{\cal Y}]$ represents a good approximation of $[{\cal X}]$ is based on the  {\it Jaccard index} \cite{FH16} $J_2({\cal X},{\cal Y})=a_{11}/\left(a_{11}+a_{10}+a_{01}\right)$. Here $a _{11}$ indicates the number of pairs of elements which are in the same cluster in both partitions, $a_{01}$ ($a_{10}$) the number of pairs of elements which are in the same cluster in $[{\cal X}]$  ($[{\cal Y}]$) and in different clusters in $[{\cal Y}]$  ($[{\cal X}]$).
The optimal partition corresponds to a maximum of $J_2({\cal X},{\cal Y})$ and we will show that it coincides with the one leading to a maximum of $z_2\left( n_{link}\right)$. 

Furthermore, in order to identify the partition $[{\cal Y}]$ with the best discrimination between immigrants and triggered elements we introduce a generalization of the  {\it Jaccard index} $J_1({\cal X},{\cal Y})=b_{11}/\left(b_{11}+b_{10}+b_{01}\right)$. Here $b_{11}$ represents the number of common immigrants in the two partitions, $b_{01}$ is the number of elements wrongly identified as immigrants in the partition $[{\cal Y}]$, whereas $b_{10}$ corresponds to the number of true immigrants identified as triggered elements in the partition $[{\cal Y}]$. We will show that the maximum of $J_1({\cal X},{\cal Y})$ corresponds to a minimum of  $z_1\left( n_{link}\right)$. We remark that $J_2({\cal X},{\cal Y})$ and $J_1({\cal X},{\cal Y})$ can only be used for validation purposes since for their evaluation the knowledge of the a-priori partition  $[{\cal X}]$ is required. Conversely, $z_1\left( n_{link}\right)$ and $z_2\left( n_{link}\right)$ are extracted directly from data.

\section{Numerical results}

We consider two models with different expressions of $p_{ij}$ in Eq.(\ref{lambda}). The first is a simplified model where the descendants occur simultaneously, neglecting in this way the temporal factor.
The second is the Epidemic Type Aftershock Sequence (ETAS) model~\cite{Oga85,Oga88,Oga88b} introduced to describe the strong spatio-temporal clustering of seismic occurrence \cite{dAGGL16}. Earthquakes, indeed, occur mainly close to large earthquakes both in time and space, a feature which is well captured by an epidemic description. 
The ETAS model is  considered the standard baseline for testing hypotheses associated with earthquake clusters \cite{Z0VJ04} and
is  widely adopted in operational forecasting by national agencies for seismic hazard. For both models we construct an a posteriori partition $[{\cal Y}\left(n_{link}\right)]$ by imposing the condition that two elements $i$ and $j$ belong to the same cluster ${\cal Y}_{k(i)}={\cal Y}_{k(j)}$ if $p_{ij}$ is larger than a reference threshold value $p_{th}$. We start from $p_{th}=0$ where $n_{link}=N\times (N-1)/2$ and increase $p_{th}$ leading to a monotonous decrease of $n_{link}$. In this way, we obtain different partitions $[{\cal Y}\left(n_{link}\right)]$ to test if $z_i\left (n_{link}\right )$ can be used to provide the information present in $J_i({\cal X},{\cal Y}(n_{link}))$, for $i=1,2$.
  
We start by considering a simple epidemic model (Fig.\ref{fig1}a,b) where $K$ immigrants are distributed in the two-dimensional space according to a random walk,  with the distance between any two immigrants uniformly distributed in $[0.5D,1.5D]$, where $D$ is a key parameter of the model. The $i$-th immigrant has $l_i\ge 1$ or $0$ descendants with probability $1/L$ and  $1-1/L$, respectively, where $l_i$ is extracted from a Gaussian distribution with mean $\overline{l}$ and standard deviation $0.1\overline{l}$. The $l_i$ descendants are isotropically distributed in space at a distance $d_{ij}$ from their immigrant  according to an exponential distribution $p_{ij}=\exp{\left(-\frac{d_{ij}}{d_0}\right)}/d_0$.
We perform simulations with $K=L=\overline{l}=5000$, obtaining $N=18984$ elements with $4980$ clusters containing only one element and we consider different values of $D/d_0$.
In Fig.\ref{fig1}c we plot $\delta {\cal LL}\left(n_{link}\right)$ and $R\left(n_{link}\right)$ as a function of $n_{link}$. We observe that for all $D/d_0$ values, $\delta {\cal LL}$ presents an initial fast decay for $n_{link} \lesssim 2E5$, an intermediate slow decay up to $n_{link} \sim 8E6$ followed by a much faster decay to its asymptotic value. The larger $D/d_0$ the sharper is the transition between the different regimes. 
The function $R\left(n_{link}\right)$ shows a similar behavior besides the lack of the asymptotic regime. According to the above considerations we expect that $n_1^*$ is located within the intermediate regime, where  $R\left(n_{link}\right)$ exhibits a slow decay, whereas $n_2^*$ is located after the fast decrease of $\delta {\cal LL}\left(n_{link}\right)$ but before converging to the asymptotic value.
This is confirmed by the comparison between $J_i({\cal X},{\cal Y}(n_{link}))$ and $z_i\left(n_{link}\right)$ for $i=1$ in panels (d) and (e) and for $i=2$ in panels (f) and (g).
We find that $z_1\left(n_{link}\right)$ exhibits a non-monotonic behavior for all $D/d_0$ values and presents its minimum  in a range of $n_{link}$ values which roughly corresponds to the interval where  $J_1({\cal X},{\cal Y}(n_{link}))$ presents its maximum. In addition, we observe that as $D/d_0$ takes larger values, $J_1$ is increasing and conversely $z_1$ is decreasing, presumably due to less overlapping among the clusters.     
Fig.\ref{fig1}g shows that $z_2\left(n_{link}\right)$ presents a sharp maximum at $n_{link} \simeq 9.0E6$.  According to Eq.(\ref{zeta}) the position of the abrupt decay of $z_2$ corresponds to the optimal threshold $ n_{link}=n_2^*$ which is therefore located on the right of  its maximum value.
More precisely, we define as $n_2^*$ the point at which $z_2\left(n_{link}\right)$ reaches the half of its peak value. Comparing with panel (f) we find that this value well corresponds to the position of the maximum of $J_2({\cal X},{\cal Y}(n_{link}))$, with a similar dependence  on $D/d_0$ values.  
These results strongly support our conjecture that the information contained in  $J_i({\cal X},{\cal Y}(n_{link}))$ can be extracted directly from $z_i\left(n_{link}\right)$, for $i=1,2$, without any a-priori knowledge of the optimal partition $[{\cal X}]$.

Next we consider the ETAS model
where the proximity between two elements $p_{ij}$ represents the probability a previous earthquake $j$, with magnitude $m_j \ge m_0$, to trigger a subsequent one $i$ and is given by 
\begin{equation}
  p_{ij} \propto
 10^{\alpha (m_j-m_0)}
 \left (1+\frac{t_i-t_j}{c}\right )^{-p_0}
 \left (1+\frac{d_{ij}^2}{D^2}\right)^{-q_0} 10^{-b(m_j-m_0)}
\label{etasQ}
\end{equation}
where $d_{ij}$ is the distance between the epicenters of the two earthquakes and $D=d_0 10^{\gamma (m_j-m_0)}$. The model contains $7$ parameters $c,p_0,d_0,\gamma,q_0,\alpha,b$ which are usually estimated via log-likelihood maximization procedures \cite{Oga83,VS08,BLGDA11,Scho13,LGdAMG14}.
Eq.(\ref{etasQ}) implements the experimental observation that the number of descendant earthquakes, usually termed aftershocks, is a power law decreasing function of the temporal and spatial distance from the triggering earthquake and exponentially depends on its magnitude.
The model can be adapted to describe the spreading of a virus if we replace the spatial distance $d_{ij}$ with a metric that quantifies the connection between two elements \cite{CWHLXM20,KSM20}.
Within this context, the dependence on the magnitude of the triggering element can model the important role played by super-spreaders in the COVID-19 pandemic \cite{Lax20}.

\begin{figure*}
  \includegraphics[width=16cm]{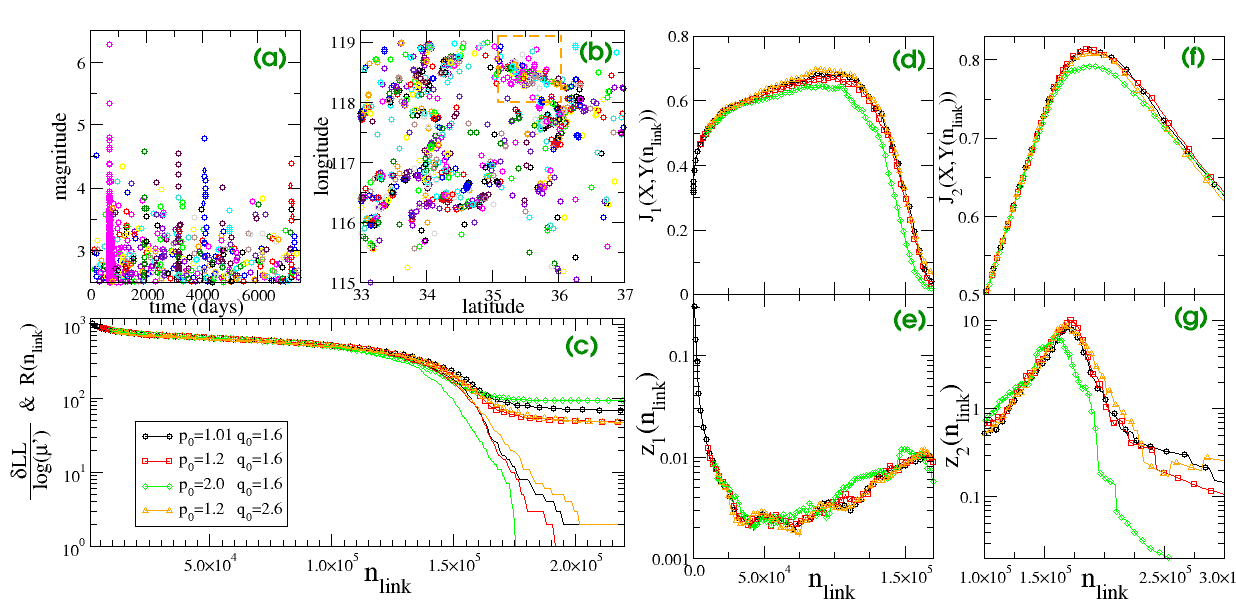}
  \caption{The magnitude of elements as a function of time for the ETAS  model (a). We adopt the numerical procedure of ref. \cite{dAGL18} with parameters optimized for the Southern California Region in \cite{PL20}, $c=0.01$ days, $p=1.2$, $d_0=0.006$ degrees,$\gamma=0.4$, $q=1.5$, $\alpha=1$ and $b=1.1$. We use the same colors and symbols for elements belonging to the same cluster in order to enlighten the temporal clustering. (b) The spatial organization of elements in the ETAS model, adopting the same color code of panel (a). The analysis of the clustering structure is focused on events inside the dashed orange rectangle. (c) Continuous lines represent $R\left(n_{link}\right)$ whereas different symbols are used for  $\delta{\cal LL}\left(n_{link}\right)$ for different values of $p_0$ and $q_0$ (see caption) in the proximity matrix $p_{ij}$ of the ETAS model (Eq.(\ref{etasQ})). In particular we use, green diamonds for $p_0=2.0,\ q_0=1.6$, red squares for $p_0=1.2,\ q_0=1.6$, orange triangles for $p_0=1.2,\ q_0=1.6$  and black circles for $p_0=1.01,\ q_0=1.6$. The same color codes and symbols are used for the index $J_1({\cal X},{\cal Y}(n_{link}))$ (d) the slope $z_1(n_{link})$ (e) the Jaccard index  $J_2({\cal X},{\cal Y}(n_{link}))$ (f) and the effective exponent $z_2(n_{link})$ (g).}   
  \label{fig2}
\end{figure*}

In Fig. \ref{fig2}a,b we plot a typical seismic pattern of 9820 earthquakes with $m\ge2.5$, simulated by the ETAS model with parameters listed in the figure caption, leading to an optimal overlap with real seismic data for Southern California. 
Fig.\ref{fig2} clearly enlightens the clustering of seismicity both in time and space around large seismic earthquakes. We restrict the study of the clustering structure inside the dashed rectangle in Fig.\ref{fig2}b which contains $N=1289$ earthquakes with $m\ge2.5$ grouped in $K=408$ clusters, that correspond to the a-priori partition $[{\cal X}]$. Most of these clusters ($336$) contain just one element whereas the cluster with the largest earthquake ($m=6.28$) contains $614$ elements. Fig.\ref{fig2}d,f show that both $J_1({\cal X},{\cal Y}(n_{link}))$ and $J_2({\cal X},{\cal Y}(n_{link}))$ are non-monotonic functions with corresponding peak values $n_1^*\simeq 9E4$ and $n_2^*\simeq 1.8E5$. The quantity  $z_1\left(n_{link}\right)$ fluctuates around  $z_1 \sim 2E-3$ for $n_{link} \in (4E4,9E4)$ reaching its minimum value for $n_{link} \simeq 7E4$. This leads to an estimate of $n_1^*$ which is slightly smaller than the one obtained from $J_1({\cal X},{\cal Y}(n_{link}))$ but falls within the interval where $J_1({\cal X},{\cal Y}(n_{link}))$ presents its plateau. In particular, for $n_{link}= 7E4$, $J_1({\cal X},{\cal Y}(n_{link)})$ takes a value which is only $5\%$ smaller than its maximum value, therefore the estimate of $n_1^*$ provided by $z_1\left(n_{link}\right)$ can be still considered sufficiently accurate.
Concerning $n_2^*$ we find that $z_2\left(n_{link}\right)$ presents a maximum at $n_{link}\simeq 1.7E5$ (Fig.\ref{fig2}g) which is slightly before the position of the maximum of $J_1({\cal X},{\cal Y}(n_{link}))$. Again, we obtain an accurate estimate of $n_2^*$ by defining it at the half-peak value of $z_2$.

In Fig.\ref{fig2} we also explore the efficiency of the method when the exact proximity matrix $p_{ij}$ is not known. More precisely, we perform the same analysis but considering in Eq.(\ref{etasQ}) values of
$p_0$ and $q_0$ which do not correspond to the ones used in the numerical generation of the catalog. We find that  $J_i({\cal X},{\cal Y})(n_{link})$, for $i=1,2$, weakly depends on the value of these parameters (Fig.\ref{fig2}d,f), apart from the largest $p_0$ value ($p_0=2$) where we find a significant reduction of $J_i({\cal X},{\cal Y}(n_{link}))$ indicating  a less accurate partitioning $[{\cal Y}]$.
The same considerations apply to $z_i\left(n_{link}\right)$, for $i=1,2$, as clearly shown in Fig \ref{fig2}e,g, meaning that the method is very efficient even if one does not implement the exact value of $p_{ij}$ in the partitioning.  
For the sake of completeness, in Fig.\ref{fig2}c we show that the dependence of $\delta {\cal LL}\left(n_{link}\right)$ and $R\left(n_{link}\right)$ on $n_{link}$ is consistent with the expected pattern.

\section{Conclusions}
We have introduced two criteria for cluster organization in stochastic epidemic models, that can be easily estimated from the data. In particular, we have shown that the optimal parameters for cluster identification can be extracted from the quantities $z_i\left(n_{link}\right)$, for $i=1,2$,  without any a-priori knowledge of the clustering structure. The quantity $z_1$ can be used for an accurate separation of immigrants and descendants, specifically for the detection of the number of clusters or communities, and $z_2$ allows us to achieve the best approximation of the internal structure of each cluster or community. We have verified the accuracy of our method, by means of a comparison with the $Jaccard\ index$, in numerical simulations of two models with different properties. In both models and for a wide range of parameters we show that the method provides the optimal identification of the clustering structure. 
We remark that the time complexity of our method is linear with the total number of links ${\cal O}(M)$ whereas in commonly adopted algorithms for community detection, as the Girvan-Newman one \cite{GN02}, time complexity in the worst case is ${\cal O}(M^2 N)$. Our algorithm therefore appears very appropriate in epidemic data sets with a large number of elements. Furthermore it can be easily implemented in automated algorithms for cluster identification and can be very efficient in the non-parametric evaluation of the proximity matrix $Q(\cal Y)$ \cite{ML08}.

\section{Acknowledgments}
E. Lippiello, acknowledges support from project PRIN201798CZLJ and from VALERE project of the University of Campania ``L. Vanvitelli''. This research is co-financed by Greece and the European Union (European Social Fund-ESF) through the Operational Programme <<Human Resources Development, Education and Lifelong Learning>> in the context of the project ``Strengthening Human Resources Research Potential via Doctorate Research'' (MIS-5000432), implemented by the State Scholarships Foundation (IKY).


\end{document}